\documentclass[12pt]{amsart}
\usepackage{amsmath}
\usepackage{amssymb}
\usepackage[latin1]{inputenc}
\usepackage[T1]{fontenc}
\newtheorem{proposition}{Proposition}

\newcommand{\tr}{{\rm Tr }}

\newcommand{\bra}{\langle}
\newcommand{\ket}{\rangle}
\newcommand{\vp}{\varphi}

\newcommand{\C}{\mathbb{C}}
\newcommand{\N}{\mathbb{N}}

\newcommand{\R}{\mathbb{R}}
\newcommand{\be}{\begin{equation}}
\newcommand{\eeq}{\end{equation}}
\newcommand{\bet}{\begin{equation*}}
\newcommand{\eeqt}{\end{equation*}}
\newcommand{\bea}{\begin{eqnarray}}
\newcommand{\eeqa}{\end{eqnarray}}
\newcommand{\beat}{\begin{eqnarray*}}
\newcommand{\eeqat}{\end{eqnarray*}}

\newcommand{\h}[1]{\mathcal{#1}}
\newcommand{\hil}{\mathcal{H}}

\begin{document}
\title{Quantization and noiseless measurements}
\author{J. Kiukas}
\address{Jukka Kiukas,
Department of Physics, University of Turku,
FIN-20014 Turku, Finland}
\email{jukka.kiukas@utu.fi}
\author{P. Lahti}
\address{Pekka Lahti,
Department of Physics, University of Turku,
FIN-20014 Turku, Finland}
\email{pekka.lahti@utu.fi}
\begin{abstract}
In accordance with the fact that quantum measurements are described in terms of positive operator measures (POMs),
we consider certain aspects of a quantization scheme in which a classical variable $f:\R^2\to \R$ is associated with
a unique positive operator measure (POM) $E^f$, which is not necessarily projection valued.
The motivation for such a scheme comes from the well-known fact
that due to the noise in a quantum measurement, the resulting outcome distribution is given by a POM and cannot, in general, be described
in terms of a traditional observable, a selfadjoint operator. Accordingly, we notice that the
noiseless measurements are the ones which are determined by a selfadjoint operator.
The POM $E^f$ in our quantization is defined through its moment operators,
which are required to be of the form $\Gamma(f^k)$, $k\in \N$, with $\Gamma$ a fixed map from classical variables to
Hilbert space operators. In particular, we consider the quantization of classical \emph{questions}, that is,
functions $f:\R^2\to\R$ taking only values $0$ and $1$. We compare two concrete realizations of the map $\Gamma$
in view of their ability to produce noiseless measurements: one being the Weyl map, and the other defined by
using phase space probability distributions.
\end{abstract}
\maketitle

\section{Introduction}

Quantization is a procedure which turns the classical description of a physical system into its quantum description.
Given the phase space of the classical description and the Hilbert space of the corresponding quantum description 
one often considers only quantization of the classical dynamical variables.

If the set of dynamical variables  corresponding to a given phase space is denoted by $\h F$,
and the set of quantum observables associated with a given Hilbert space by $\h E$, a \emph{quantization} (of observables)
is thus a method which assigns to
any $f$ in some subset $\h Q$ of $\h F $,  a unique $E^f\in \h E$.
The subset $\h Q$ depends on the technical implementation of the quantization scheme; it consists of those variables for which
the procedure can be applied. For each $f\in \h Q$, the observable $E^f\in \h E$ obtained by the procedure is called the \emph{quantization}
of the classical variable $f$.

Quantum observables are traditionally represented as selfadjoint operators, or, equivalently, spectral measures.
Therefore, the existing quantization schemes, like
the Weyl quantization, produce operators as quantizations. In modern quantum mechanics, however, a normalized
positive operator  measure (also called semispectral measure or  POM, for short) is needed to represent an observable.
Hence, adopting the modern definition for a
quantum observable, we face the fact that the existing quantization schemes are not conceptually satisfactory,
as they produce only a limited class of observables. 
It is well-known, that the outcome statistics of a typical quantum measurement cannot, in general,
be described by a single selfajoint operator.

The purpose of this note is to describe and further elaborate the quantization scheme introduced in \cite{quantization}.
This scheme produces observables (as POMs) which are not determined by single selfadjoint operators but by sequences of
operators. In order to simplify the presentation, we consider only a particle moving in one dimension
so that the phase space is $\R^2$ and the Hilbert space is $L^2(\R)$.
We mention, however, that
the scheme was defined in \cite{quantization} for a more general case where the phase space is a locally compact
group.

\section{The quantization method}

The phase space points $(q,p)\in \R^2$ represent the (pure) states of the
classical system, 
and its dynamical variables are given 
as (Borel) functions $f:\R^2\to \R$. Let $\h F(\R^2)$ denote the set of all these variables.
The corresponding quantum system is attached by the Hilbert space
$\hil=L^2(\R)$, 
the unit vectors $\vp\in\hil$ representing the (pure) states of the system and its
observables being given by POMs $E:\h B(\R)\to L(\hil)$. Here $\h B(\R)$ denotes the $\sigma$-algebra of the Borel sets of $\R$ and
$L(\hil)$ is the set of bounded operators on $\hil$.

Let $\Omega$ stand either for $\R$, or $\R^2$, and let $\h A$ be the corresponding
$\sigma$-algebra, that is, $\h B(\R)$, or $\h B(\R^2)$.
If $E:\h A\to L(\hil)$ is a POM and $f:\Omega \to \C$ a (measurable) function, we let
$\int f dE = \int f(\omega)dE(\omega)$
denote the operator integral of $f$ with respect to $E$ according to the theory given in \cite{Lahti}. 
For subsequent use, we need to recall the domain $D(f,E)$ of the operator $\int f dE$.
To this end, for any $\psi,\vp\in\hil$, we let $E_{\psi,\vp}$ denote the complex measure $X\mapsto \langle\psi|E(X)\vp\rangle$.
Then
\beat
D(f,E)&=&\{ \vp\in\hil \mid f \text{ is } E_{\psi,\vp}\text{-integrable for each } \psi\in\hil \}\\
&\supset& \{  \vp\in \hil \mid \int |f|^2 dE_{\vp,\vp}<\infty\} :=\widetilde{D}(f,E). 
\eeqat

Let $\h E(\hil)$ denote the set of all observables of the quantum system, i.e the set of  POMs $E:\h B(\R)\to L(\hil)$.
For each observable $E$, we can define the \emph{moment operators} $E[k]$, $k\in \N$, of $E$ via
\bet
E[k] = \int x^k dE(x).
\eeqt
It should be noted that
these  operators are not necessarily densely defined.

Finally, let $\h O(\hil)$ be the set of all (not necessarily bounded or even densely defined) linear operators in $\hil$.
For each $A\in \h O(\hil)$ we denote by $D(A)$ the domain of  $A$.
The symbols $Q$ and $P$ stand for the standard position and momentum operators, i.e. $(Q\vp)(x)= x\vp(x)$ and
$(P\vp)(x)=-i\frac{d}{dx}\vp(x)$, with their usual domains, and the spectral measures $E^Q$ and $E^P$, respectively. 
Also, let $[O,I]\subset L(\hil)$ be the set of \emph{effects},
i.e. operators $A\in L(\hil)$ with $0\leq A\leq I$.

\subsection{The quantization}\label{secquant}

First of all, we assume that we have a fixed map $\Gamma:\h U_{\Gamma} \to \h O(\hil)$, where $\h U_{\Gamma}$ is a subset of $\h F$.
\emph{We want to emphasize that, contrary to the traditional schemes, the operators $\Gamma(f)$ are {\bf not} regarded as the
quantizations of the classicals variables $f\in \h U_{\Gamma}$.}

Let $k\in \N$, and consider a classical variable $f\in \h F$. Then the number
$\int x^k \mu_{(q,p)}(f^{-1}(dx))= f^k(q,p)$, where $\mu_{(q,p)}$ denotes the point measure at $(q,p)\in \R^2$, gives
the $k$th moment of the measurement outcome distribution of the dynamical variable $f$ 
in the  state $(q,p)$. 
Similarly,  the integral $\int x^k dE_{\vp,\vp}= \langle\vp|E[k]\vp\rangle$
is the $k$th moment of the measurement outcome distribution $E_{\vp,\vp}$ of the quantum 
observable $E\in \h E(\hil)$ in the  state $\vp$. (Of course, the vector $\vp$ must be in the domain of the operator
$E[k]$ for this to be valid.)

Now, we want to define a quantization $E^f\in \h E(\hil)$ of a
classical variable $f\in \h U_{\Gamma}$ by using information on the classical moments $f^k$, in such a way that the
fixed map $\Gamma$ transfers these moments to the moments of the observable $E^f$, namely 
$\langle\vp|\Gamma(f^k)\vp\rangle=\langle\vp|E^f[k]\vp\rangle$
for each  $\vp$ (with $\vp$ belonging to the common domain of the operators).
Of course, this happens if we require that each $f^k$ belongs to $\h U_{\Gamma}$ and
\be\label{moments}
\Gamma(f^k)= E^f[k], \ \ \ k\in \N.
\eeq
Hence, in order to apply this procedure, we have to solve an operator {\em moment problem}.
In addition, as mentioned in the Introduction,
a quantization procedure should always lead to a unique $E^f$. This means that a given solution
to the moment problem has to be \emph{determinate}, i.e. no other POM  $\h B(\R)\to L(\hil)$ may
have the same moment sequence. We let $\h Q_{\Gamma}\subset \h U_{\Gamma}$ denote the set of variables
for which these requirements are satisfied.

Hence, \emph{our quantization is the association $\h Q_{\Gamma}\ni f\mapsto E^f\in \h E(\hil)$}, instead of the
map $\h U_{\Gamma}\ni f\mapsto \Gamma(f)\in \h O(\hil)$. Notice, in particular, that for a classical variable $f\in \h Q_{\Gamma}$,
the operator $\Gamma(f)$ does not, in general, determine the quantization $E^f$.

Naturally, for different maps $\Gamma$, the sets $\h Q_{\Gamma}$ may be very different, in particular,
$\h Q_{\Gamma}$ can easily be empty, in which case the map is, of course, quite useless. For a (trivial) example,
consider the case where $\Gamma$ is defined by $\Gamma(f)=A$ for all $f\in \h F$, with  $A\in \h O(\hil)$  fixed.
If $A$ is bounded but not positive, then \eqref{moments} is not satisfied by any $f\in \h F$ and  $E\in\h E(\hil)$, because
for some $\vp\in \hil$ and any choice for $f$ and $E$ satisfying \eqref{moments}, we have
$\int x^2 dE_{\vp,\vp} = \bra \vp |A\vp\ket<0$, a contradiction. Hence $\h Q_{\Gamma}$ is empty in this case.

For a given map $\Gamma$, the natural task would be to investigate which variables can be quantized by the ensuing scheme,
i.e. which functions constitute the set $\h Q_{\Gamma}$. Since this is obviously
a very difficult problem, we content ourselves with some simple, yet important functions.

In \cite{quantization}, we demonstrated that for certain choices of the map $\Gamma$, the associated quantization scheme
actually produces meaningful results. Namely, in these cases the classical position and momentum variables
$x:(q,p)\mapsto q$ and $y:(q,p)\mapsto p$ indeed belong to $\h Q_{\Gamma}$, and the ensuing operators $\Gamma(x^k)$ and $\Gamma(y^k)$
can be determined. These operators turn
out to be 
polynomials of the position and momentum operators $Q$ and $P$, respectively (\cite[Theorem 4]{quantization}), and the
corresponding observables are unsharp  position and momentum observables; neither of them can be represented
by  a single selfadjoint operator. It was proven in \cite{Dvurecenski} that the moment problem of this example is indeed determinate.
The realization of the map $\Gamma$ used is given in Section~\ref{implementationseca}.

\subsection{Quantization of question variables}

For any variable $f\in \h F$, one can consider the family of
variables $\chi_B\circ f$, where $B$ goes through the Borel sets of the real line (and $\chi_B$ denotes the indicator
function of the set $B$). Such variables are traditionally called \emph{questions} \cite{Mackey}, since $\chi_B\circ f$
formalizes the question of whether the value of $f$ lies in the set $B$. Since each $f\in \h F(\R^2)$ is 
determined by the totality of such questions, one could try to quantize these questions instead of
the variable itself.  It is then another problem
whether the resulting family of simple observables can constitute  
a single observable describing the quantization of the original variable $f$.

In any case, this approach leads us to consider the functions of the form $\chi_X$, with $X\in \h B(\R^2)$, since
all possible questions are of this form. Now the moment problem becomes very simple,
as the sequence in the relevant operator moment problem is the constant sequence $(\Gamma(\chi_X^k))_{k\in \N}$.

Hence, we have the (seemingly trivial) moment problem
\be\label{constantmoment}
E[k]=A, \ \ \ k\in \N,
\eeq
where $A\in \h O(\hil)$ is fixed, and $E\in \h E(\hil)$ is a solution. For a $A\in [O,I]$,
let $E^A:\h B(\R)\to L(\hil)$ be the  two-valued  POM 
supported in the  set $\{0,1\}$ with $E^A(\{0\})= I-A$ and $E^A(\{1\})= A$. It is clear that $E^A$ is a solution for
\eqref{constantmoment}. The following observation is also immediate.

\begin{proposition}
Let $A\in \h O(\hil)$ be such that $D(A)$ is dense in $\hil$. Then the moment problem \eqref{constantmoment} has a solution
if and only if $A\in [O,I]$. In that case, the only solution is the operator measure $E^A$.
\end{proposition}
\begin{proof} Let $A\in \h O(\hil)$ be densely defined. We are left to prove that for a densely defined $A\in \h O(\hil)$,
the existence of a solution $E$ for \eqref{constantmoment} forces $A$ to be bounded (with $D(A)=\hil$) and $0\leq A\le I$, and
$E$ to be $E^A$. To that end, assume that $A\in \h O(\hil)$ with $D(A)$ dense, and that
$E:\h B(\R)\to L(\hil)$ is a  POM satisfying \eqref{constantmoment}.
Then $D(A)=D(x^{2k}, E)\subset \widetilde{D}(x^k,E)\subset D(x^k,E)=D(A)$ for all $k\in \N$, so $D(A)=D(x^{2k}, E)=\widetilde{D}(x^k,E)$
for all $k\in \N$. Given $\vp\in D(A)$, $\|\vp\|=1$, we have $0\leq \int x^2 dE_{\vp,\vp} =\bra \vp |A\vp\ket$, and
\bet
0\leq \int (x-\bra \vp|A\vp\ket)^2 dE_{\vp,\vp} = \bra \vp|A\vp\ket-\bra \vp|A\vp\ket^2,
\eeqt
so that $0\leq \bra \vp|A\vp\ket \leq 1$, and we can define a discrete probability measure $\mu_{\vp}$ by the rules
$\{0\}\mapsto 1-\bra \vp|A\vp\ket$ and $\{1\}\mapsto \bra \vp|A\vp\ket$. Now the measures
$E_{\vp,\vp}$ and $\mu_{\vp}$ have the same (constant) moment sequence for all $\vp\in D(A)$. This is possible
only if $\mu_{\vp}=E_{\vp,\vp}$ (see e.g. \cite[Theorem 2.2]{Freud} and the Remark belonging to it).
It follows that $\bra \vp|E(\{0\})\vp\ket + \bra \vp|E(\{1\})\vp\ket = 1$, for all $\vp\in D(A)$,
$\|\vp\|=1$. Since $D(A)$ is dense, this implies $E(\{0\})+E(\{1\})=I$, i.e. the operator measure $E$ is supported
in $\{0,1\}$. But then $D(A)=D(x,E)=\hil$. In addition, $\bra \vp |E(\{1\})\vp\ket = \bra \vp|A\vp\ket$ for $\vp\in D(A)$
so that $A=E(\{1\})$. Hence, $A$ is bounded, with $D(A)=\hil$ and $0\leq A\leq I$.
Moreover, we have $E=E^A$, so that the proof is complete.
\end{proof}
\noindent {\bf Remark.} Note that the moment problem \eqref{constantmoment} may have multiple solutions, if $D(A)$ is not dense.
This is easily demonstrated by a simple example: Fix $a\in \R$, $0\leq a\leq 1$, and a nontrivial proper closed subspace
$\h M$ of $\hil$. Let $A=aI|_{\h M}$, so that $A\in \h O(\hil)$, with $D(A)=\h M$. Let $\mu$ be the probability measure with
$\{0\}\mapsto 1-a$ and $\{1\}\mapsto a$. For any probability measure $\nu:\h B(\R)\to [0,1]$, define a  POM $E^{\nu}$
by $E^\nu (B) = \mu(B)P + \nu(B)(I-P)$, $B\in \h B(\R)$, where $P$ denotes the projection onto $\h M$. Now if the measure
$\nu$ is chosen such that $\int_{\R} |x|d\nu(x)=\infty$, it is easy to verify that $D(x^k,E^{\nu})=\widetilde{D}(x^k,E^{\nu})=\h M$,
and
\bet
\bra \psi|E^{\nu}[k]\vp\ket = \int x^k d\mu \bra \psi|\vp\ket = a \bra \psi|\vp\ket,  \ \ \vp\in \h M, \  \psi\in \hil, \ k\in \N,
\eeqt
so that $E^{\nu}[k]= A$ for all $k\in \N$. Hence, the moment problem \eqref{constantmoment} has
(in fact, uncountably) many different solutions in this case.

\

The preceding proposition suggests that in order to be able to quantize at least all the questions, the quantization 
map $\Gamma$  should be chosen  such  that $\chi_X\in \h U_{\Gamma}$ and $\Gamma(\chi_X)\in [O,I]$ for all
$X\in \h B(\R^2)$. We note that, together
with linearity and certain simple convergence assertions, this condition actually forces the map $X\mapsto \Gamma(\chi_X)$
to coincide with a  POM $G:\h B(\R^2)\to L(\hil)$, and consequently, to have the property that
$\Gamma(f)= \int f dG$ for each bounded function $f\in \h U_{\Gamma}$ (see e.g. \cite{Werner, NCQM, quantization}).
Indeed, the realization  discussed in Section~\ref{implementationseca}. is of this type.

It should be emphasized that the selfadjoint operator $\Gamma(\chi_X)\in [O,I]$ is \emph{not} the
observable representing the quantization of the question $\chi_X$, since its spectral measure does not satisfy
\eqref{constantmoment} with $A=\Gamma(\chi_X)$, except in the case where $\Gamma(\chi_X)$ is a projection.
(As mentioned in the proposition, the only POM satisfying \eqref{constantmoment} is the discrete POM $E^A$.) 

However, we do not require that $\Gamma$ should map all questions to their traditional
quantum mechanical counterparts, i.e. projections. This is essential, since otherwise the POM $G:\h B(\R^2)\to L(\hil)$
in the above mentioned important case $\Gamma(f) = \int f dG$ would be projection valued, and
consequently 
all the quantized observables would be given as mutually commuting spectral measures, which cannot be the case.
Actually, even the presence of a single nontrivial projection in the range of $G$ would mean that the ensuing quantum system
of observables would have a nontrivial classical property. In fact, if $\Gamma(\chi_X)=G(X)$ is a nontrivial
($\ne O,I$) projection for some $X\in \h B(\R^2)$, then this projection commutes with any of the quantized observables \cite{Stulpe},
which means that the quantization $\Gamma$ leads to a superselection rule for the resulting quantum system
\cite{Jauch,BeltramettiCassinelli}.

\subsection{Quantization and noiseless measurements}
In quantum mechanics a measurement always determines an observable $E\in\h E(\hil)$; in fact, quantum observables
can be viewed as equivalent
classes of measurements, see, e.g. \cite{Holevo, HolevoII,QTM}. The {\em noise operator} $N(E)=E[2]-E[1]^2$ of an observable $E$ can be
taken to describe the inherent  inaccuracy in a measurement of $E$.
Indeed, for any  state $\varphi$ (for which the quantities involved are well defined) the variance ${\rm Var}(E,\varphi)$
(square of the standard deviation) of the measurement outcome statistics of $E$ can be written as 
$$
{\rm Var}(E,\varphi) = {\rm Var}(E[1],\varphi)+\langle\vp|N(E)\vp\rangle,
$$
where ${\rm Var}(E[1],\varphi)$ is the variance of the corresponding outcome statistics of the spectral measure of
$E[1]$, which is assumed to be selfadjoint.
We say that a measurement is {\em noiseless} if the noise operator of the associated observable $E$ is  zero.

As a measure for the degree of unsharpness of the measurement, the noise operator also
indicates to what extent the determined observable $E$ fails to be sharp, i.e. a spectral measure.
Accordingly, we know from \cite[Theorem]{quantization} that, in the case where $E[1]$ happens to be selfadjoint, the measurement is
noiseless exactly when $E$ is a spectral measure.

Consider now a quantization map $\Gamma$ and assume that it is given by a POM $G:\h B(\R^2)\to L(\hil)$.
Let $E^f$ be a quantization of a classical variable $f$ by $\Gamma$ and assume that it is projection valued. Then any
projection $E^f(B)$, $B\in\h B(\R)$, would commute with any other quantized observable $E^g$, $g\in\h O_\Gamma$, and, therefore,
$E^f$ would constitute a superselection rule for the quantum description, as noted above.
Hence, if the quantized system of observables is a proper quantum system, with no
nontrivial classical properties, then the quantized system admits no noiseless measurements. 

\section{Two realizations for the map $\Gamma$}\label{implementationsec}

In order to produce an applicable (and meaningful) quantization scheme, the map $\Gamma$ has to be chosen properly.
Rather that trying to investigate general conditions which could be required for such a $\Gamma$,
we consider, and compare, two specific ways to attempt to define it.

Let $W(q,p)$ be the Weyl operators; $W(q,p)= e^{\frac 12iqp} e^{-iqP}e^{ipQ}$. If $A\in L(\hil)$, define
$\Gamma^A:\h F(\R^2) \to \h O(\hil)$ (formally) by
\be\label{gamma}
\Gamma^A(f) = \frac{1}{2\pi} \int_{\R^2} f(q,p) \ W(q,p)AW(q,p)^* dqdp, \ \ \  f\in \h F(\R^2).
\eeq
This definition can be made precise in different ways. We consider the following two types.

\subsection{Type (a)}\label{implementationseca}

Let $A$ be positive, with $\tr[A]=1$. We will call such an operator a \emph{generating operator}, and denote it by $T$.
Choose  $\h U_{\Gamma}=\h F$. For any $f\in \h F$, we define $\Gamma^T(f)$ as the operator determined by 
\bet
\bra \psi|\Gamma^T(f)\vp\ket = \frac{1}{2\pi} \int_{\R^2} f(q,p) \bra \psi|W(q,p)TW(q,p)^*\vp\ket dqdp,
\eeqt
with $\psi\in \hil$ and $\vp\in D(\Gamma^T(f))$, the domain $D(\Gamma^T(f))\subset \hil$ of $\Gamma^T(f)$ consisting of those vectors
$\vp\in \hil$ for which the above integral exists for all $\psi\in \hil$. It can be shown (see \cite[Proposition 2]{quantization}) that
such an operator indeed exists.
We used this realization in \cite{quantization} to quantize position and momentum variables, for a certain kind of
generating operators $T$. The resulting operators $\Gamma^T(x^k)$ and $\Gamma^T(y^k)$
were found to be densely defined and selfadjoint; in fact, they are
polynomials of degree $k$ of $Q$ and $P$, respectively.

The determination of the operators $\Gamma^T(f)$ is simplified by the fact that we can, in many cases, consider only the integrals
over the probability distributions
\bet
(q,p)\mapsto P^T_{\vp}(q,p)=\bra \vp |W(q,p)TW(q,p)^*\vp\ket.
\eeqt
In fact, the set $\widetilde{D}(\Gamma^T(f))$ of those $\vp\in \hil$ for which $f^2$ is integrable
with respect to the aforementioned density, is a subspace of the domain $D(\Gamma^T(f))$. Accordingly, some authors define the operator
integral \eqref{gamma} in that smaller set in the first place (see e.g. \cite{Wernerscreen}). We prefer, however, to use the natural
domain $D(\Gamma^T(f))$.

The map $\Gamma^T$ has the convenient property that $\Gamma^T(f)$ is a bounded operator defined in all of $\hil$,
whenever $f\in \h F(\R^2)$ is a bounded function.

Moreover, we recall that in the case where $T$ is the one-dimensional projection determined by the vector $h_0$, 
with $h_0(x)=\pi^{-\frac 14} e^{-\frac 12 x^2}$ (the ground state of the oscillator), the probability distribution 
$P^T_{\vp}$ is the {\em  Husimi distribution}.
Sometimes the associated quantization $f\mapsto \Gamma^{h_0}(f)$ is called the {\em Berezin quantization}
(see e.g. \cite{Landsman}).

\subsection{Type (b)}\label{implementationsecb}

Take $A=\h P$, the parity operator ($(\h P \vp)(x)= \vp(-x)$), in which case  the map
$\Gamma^{\h P}$ is (formally) the {\em Weyl quantization}. There are different ways to give meaning to the integral
\eqref{gamma} in this case, depending on the properties of the function $f$ (see e.g. \cite{Pool,Dubin,Werner,Daubechies}).
Most satisfactory and general meaning is given in terms of distributions \cite{Daubechies, Dubin}, but then
the resulting object $\Gamma^{\h P}(f)$ does not have to be an operator.
One could also use the same definition for \eqref{gamma} as in (a) above, since that definition works for any bounded operator
$A$ \cite[Proposition 2]{quantization}. However, usually there seems to be no way to
determine whether the domain of that operator contains
any nonzero vectors at all, except for $f\in L^1(\R^2)$ or $f\in L^2(\R^2)$, in which cases the corresponding operators
are bounded (and, in the latter case, of Hilbert-Schmidt class; see the pioneering work of Pool \cite{Pool}).

The essential cause of the difficulties in dealing with the integral \eqref{gamma} for $A=\h P$ is, of course,
the fact  the Wigner function
$(q,p)\mapsto f^W_{\vp}(q,p)=\bra \vp |W(q,p)\h P W(q,p)^*\vp\ket$ 
of a state $\vp$ is a probability density only when $\vp$ is $h_0$ (modulo phase)  \cite{Hudson}, so that we do not
have a counterpart of $\widetilde{D}(\Gamma(f))$ of (a) in this case.
Another weakness of the Weyl quantization map is that bounded functions need not lead to bounded operators (see e.g. \cite{Daubechies}).

For our (mainly illustrative) purposes, it suffices to adhere to a definition which works in some cases
(see e.g. \cite[Proposition 2]{Werner} and \cite[Proposition 8.31]{Dubin}). It is the following. First, fix $\h D\subset \hil$
to be the Schwarz space, i.e. the dense subspace consisting of infinitely differentiable functions, whose derivatives fall of faster
than any power of $x$ at infinity. Let $\h U_{\Gamma^{\h P}}$ be the set of those $f\in \h F(\R^2)$ for which there is an $R^f\in \h O(\hil)$
with the domain $\h D$, such that
\be\label{weyldef}
\int f(q,p) f^W_{\vp}(q,p) dq dp =\bra \vp|R^f\vp\ket, \ \ \ \vp\in \h D.
\eeq
Due to the assumption that $\h D$ is dense (and by  polarization), the (clearly symmetric) operator $R^f$ is uniquely determined.
Hence, we can define
$\Gamma^{\h P}(f)$ for any $f\in \h U_{\Gamma^{\h P}}$ as the closure of $R^f$.
It should be stressed that now the integral need not exist for all $\vp\in D(\Gamma^{\h P}(f))$, even if the function $f$ and the operator
$\Gamma^{\h P}(f)$ are bounded (see \cite{Werner}, the discussion after the proof of Proposition 2).

\subsection{Application to the question variables}

Consider then the quantization of the question variables  with these two types of $\Gamma$. Since each question $\chi_X$ is bounded, type (a)
gives $\Gamma^T(\chi_X)$ as a bounded operator defined in the whole $\hil$, for all generating operators $T$. Moreover, it
is well known that $0\leq \Gamma^T(\chi_X)\leq I$. Hence, according to the Proposition, each question $\chi_X$ can be quantized,
and the resulting simple observable is the POM supported in $\{0,1\}$ with $\{1\}\mapsto \Gamma^T(\chi_X)$. Since
the operator $\Gamma^T(\chi_X)$ is {\em never} a nontrivial projection (see \cite[Proposition 3]{quantization}, or
\cite[Theorem 7]{Schroeck}),
this observable is not a spectral measure. A fortiori, the totality of these simple observables constitute
a proper quantum system, and the map $\Gamma^T$ cannot produce \emph{any} observables associated with noiseless measurements.

Type (b), however, is more problematic, since the existence of the required operator $R^{f}$ is not a priori guaranteed, if the set
$X$ does not have finite Lebesgue measure. Werner has shown \cite[Proposition 2]{Werner} that if
$X$ is a sector in $\R^2$, then $R^f$ exists, and the resulting operator $\Gamma^{\h P}(\chi_X)$ is (everywhere defined)
bounded and selfadjoint. Examples of these questions $\chi_X$
are the ones of the form $\chi_B\circ f$, where $f$ is the function $f(q,p)=-\frac qp$, describing the arrival time to
the origin of a particle in place $q<0$ with the momentum $p>0$. However, in these cases, the spectrum of $\Gamma^{\h P}(\chi_X)$ is never a
subset of $[0,1]$, so according to the Proposition, our moment problem has no solution in these cases.

If $X\in \h B(\R^2)$ has a finite Lebesgue measure, then $\Gamma^{\h P}(\chi_X)$ can be defined as a bounded operator without
difficulty. However, the operator need not be positive, and so our quantization fails again. Hence, type (b) is usually
not suitable for our scheme. However, in some cases it works, as is demonstrated in the next section.

\section{Position and momentum and the associated questions}

Let $x$ and $y$ denote the classical position and momentum variables $(q,p)\mapsto q$ and $(q,p)\mapsto p$, respectively. For
$B\in \h B(\R)$, the associated questions are $\chi_B\circ x = \chi_{B\times \R}$ and $\chi_B\circ y=\chi_{\R\times B}$.
We compare the applicability of the above two types of $\Gamma$ in quantizing these basic questions.

As mentioned in the preceding section, a map $\Gamma$ of type (a) can always be applied to produce a quantization of the above
questions. In this case, type (b) can also be used. Let $F$ be the Fourier-Plancherel operator. Then, for any $\vp\in \h D$,
we have
\beat
\int_B \int f^W_{\vp}(q,p) dpdq &=& \int_B \overline{\psi(q)}\vp(q)dq,\\
\int_B \int f^W_{\vp}(q,p) dqdp &=& \int_B \overline{(F\psi)(p)}(F\vp)(p) dp,
\eeqat
as is well known.
Hence, according to the definition, $\chi_{B\times\R},\chi_{\R\times B}\in \h U_{\Gamma}$, with
$\Gamma^{\h P}(\chi_{B\times \R})=E^Q(B)$, and $\Gamma^{\h P}(\chi_{\R\times B})= E^P(B)$.

Now, since $E^Q(B)$ and $E^P(B)$ are projections, it follows by the Proposition that their respective spectral measures
$\{1\}\mapsto E^Q(B)$ and $\{1\}\mapsto E^P(B)$ 
are the quantizations of the original questions $\chi_B\circ x$ and $\chi_B\circ y$, respectively. Hence, we
see that our scheme, when implemented by using the Weyl map $\Gamma^{\h P}$, gives precisely the traditional Weyl quantizations
of the questions concerning position and momentum. In particular, Weyl quantization is eligible to produce noiseless measurements. 

In a similar way one sees that each power of the position and momentum variables themselves belong to $\h U_{\Gamma}$, and
that $\Gamma^{\h P}(x^k)=Q^k$ and $\Gamma^{\h P}(y^k)=P^k$ \cite[Proposition 8.31]{Dubin}. This means that the moment problem \eqref{moments} indeed
has a solution in both these cases, as well. The solutions are, of course, the spectral measures $E^Q$ and $E^P$, respectively.
Now a spectral measure is always determinate, i.e. no other normalized POM may have the same moment sequence. (This
follows e.g. from \cite[Theorem 5]{quantization}.) Hence, the spectral measures $E^Q$ and $E^P$ are the quantizations
of position and momentum, in the scheme based on the map $\Gamma^{\h P}$. This is, again, the  Weyl quantization.

Consider now a map $\Gamma^T$ of type (a) for some generating operator $T$. The operators $\Gamma^T(\chi_{B\times \R})$ and
$\Gamma^T(\chi_{\R\times B})$, which determine the quantizations of $\chi_{B\times \R}$ and $\chi_{\R\times B}$,
are bounded operators determined by the integrals
\beat
\bra \vp|\Gamma^T(\chi_{B\times \R})\vp\ket &=& \int_B\int P_{\vp}^T(q,p) dpdq
= \int_B\sum_n t_n (|\eta_n(-\cdot)|^2*|\vp(\cdot)|^2)(q) dq,\\
\bra \vp|\Gamma^T(\chi_{\R\times B})\vp\ket &=& \int_B\int P_{\vp}^T(q,p) dqdp
= \int_B\sum_n t_n (|(F\eta_n)(-\cdot)|^2*|(F\vp)(\cdot)|^2)(p) dq,
\eeqat
where $T=\sum_n t_n |\eta_n\ket\bra \eta_n|$. The calculation of these integrals is well known and
can be found e.g. in \cite[p. 43]{Davies}. The operators $\Gamma^T(\chi_{\R\times B})$ and $\Gamma^T(\chi_{B \times \R})$
are smeared versions of the corresponding spectral projections $E^Q(B)$ and $E^P(B)$, as can be seen by the above integral formulas, the
smearing depending on the generating operator $T$. It should be mentioned that the totality of the quantized questions
$\{E^{\Gamma^T(B\circ x)}\mid B\in \h B(\R)\}$ constitute an observable in the natural way:
$B\mapsto E^{\Gamma^T(B\circ x)}(\{1\})= \Gamma^T(B\circ x)$ is a POM, which is known to be an  \emph{unsharp} position
observable. Similarly,
$B\mapsto E^{\Gamma^T(B\circ y)}(\{1\})= \Gamma^T(B\circ y)$ is an  unsharp
momentum observable.

It was mentioned in \cite{quantization} that in the case where $T$ is the projection determined by  an oscillator eigenstate  $h_n$, $n\in\N$, we have
$x,y\in \h Q_{\Gamma^T}$, i.e. the position and momentum variables themselves can be quantized. The resulting
unique solutions to the moment problem \eqref{moments} are then precisely the above unsharp position and momentum observables.
Unfortunately, we have not been able to establish the uniqueness part of the moment problem \eqref{moments} for
a general generating operator $T$.

It is interesting to compare the two quantization methods a bit further.
In both cases one obtains the totalities of the quantized questions, 
\begin{eqnarray*}
&&\{E^{\Gamma^{\h P}(B\circ x)}\mid B\in \h B(\R)\} \ {\text{and}}\  \{E^{\Gamma^{\h P}(B\circ y)}\mid B\in \h B(\R)\},\\
&&\{E^{\Gamma^T(B\circ x)}\mid B\in \h B(\R)\}  \ {\text{and}}\  \{E^{\Gamma^T(B\circ y)}\mid B\in \h B(\R)\}.
\end{eqnarray*}
In both cases these totalities determine unique POMs; in the first case they are the spectral measures $E^Q$ and $E^P$  of $Q$ and $P$, in the second case they
constitute the unsharp position and momentum  observables $B\mapsto \Gamma^T(B\circ x)$ and $B\mapsto \Gamma^T(B\circ y)$, respectively.
In the first case $E^Q$ and $E^P$ are also the quantizations of the classical variables $x$ and $y$, whereas in the second case this
is known to be the case whenever the generating operator $T$ is given by a Gaussian $h_n, n\in\N$.
In both cases the quantized observables are totally noncommutative. In the first case they are also noncoexistent (that is, they cannot be 
measured togethre), whereas in the second case they are always coexistent (that is, they can be measured together).

\end{document}